\documentclass[aps,superscriptaddress,showpacs,preprintnumbers,superscriptaddress, nofootinbibt,twocolumn]{revtex4}
\usepackage{eurosym}
\usepackage{amsfonts}
\usepackage{amssymb,amsmath,enumitem}
\usepackage{graphicx,color,epstopdf}
\usepackage{subfigure}

\def\be{\begin{equation}}
\def\ee{\end{equation}}
\def\bea{\begin{eqnarray}}
\def\eea{\end{eqnarray}}

\begin{document}

\title{Anisotropic Dark Matter Stars}

\author{P.H.R.S. Moraes}
\email{moraes.phrs@gmail.com}
\affiliation{Universidade de S\~ao Paulo (USP), Instituto de Astronomia, Geof\'isica e Ci\^encias Atmosf\'ericas (IAG), Rua do Mat\~ao 1226, Cidade Universit\'aria, 05508-090 S\~ao Paulo, SP, Brazil} 
%\author{J.D.V. Arba\~nil}
%\affiliation{Universidad Privada del Norte, Departamento de Ciencias, Av. Alfredo Mendiola 6062 Urb. Los Olivos, Lima, Lima, Per\'u}
\author{G.~Panotopoulos}
\email{grigorios.panotopoulos@tecnico.ulisboa.pt}
\affiliation{Centro de Astrof\'isica e Gravita\c c\~ao-CENTRA, Departamento de F\'isica, Instituto Superior T\'ecnico-IST, Universidade de Lisboa-UL, Av. Rovisco Pais, 1049-001 Lisboa, Portugal}

\author{I.~Lopes}
\email{ilidio.lopes@tecnico.ulisboa.pt}
\affiliation{Centro de Astrof\'isica e Gravita\c c\~ao-CENTRA, Departamento de F\'isica, Instituto Superior T\'ecnico-IST, Universidade de Lisboa-UL, Av. Rovisco Pais, 1049-001 Lisboa, Portugal}

\begin{abstract}
The properties of exotic stars are investigated. In particular, we study
objects made entirely of dark matter and we take into account intrinsic anisotropies
which have been ignored so far. We obtain exact analytical solutions to the structure equations and we
we show that those solutions i) are well behaved within General Relativity {\it and} ii) are capable of 
describing realistic astrophysical configurations. 
\end{abstract}

\maketitle

%\tableofcontents

%%%%%%%%%%%%%%%%%%%%%%%%%%%%%%%%%%%%%%%%%
\section{Introduction}\label{sec:intro}
%%%%%%%%%%%%%%%%%%%%%%%%%%%%%%%%%%%%%%%%%

Dark matter (DM) has certainly been one of the greatest mysteries of  Physics. An important evidence of its existence came from the analysis of rotation curves of spiral galaxies by V. Rubin and collaborators in the 70's of the last century \cite{rubin/1978,rubin/1979,peterson/1978}. DM is thought to be a kind of matter that does not interact electromagnetically and therefore cannot be seen, which is why it is called {\it dark}. However, it interacts gravitationally. In the case of spiral galaxies, it causes their rotation curves to be significantly higher than one would expect by measuring only the gravitational field of luminous matter.

\smallskip
It also has a fundamental role in the formation of galaxies and large-scale structures in the universe \cite{hu/1998,blumenthal/1984,frenk/1988,gelb/1994}. Actually it is believed that when baryonic matter decoupled from radiation at redshift $z\sim1100$, the DM gravitational potential wells were already formed and rapidly attracted baryonic matter, what has speeded up the structure formation mechanism \cite{ryden/2003,dodelson/2003} so that we can see the large-scale structures we see today.

\smallskip
Moreover, according to standard cosmological model matched with observational data coming from temperature fluctuations in the cosmic microwave background radiation, DM makes up roughly $25\%$ of the matter density of  entire universe composition \cite{planck_collaboration/2016}.

\smallskip
We still have not detected DM particles with experimental apparatus despite the efforts \cite{aalbers/2016,harnik/2012,ruppin/2014,ahlen/2010}. So far we have only detected its gravitational effects when pointing telescopes to the sky. On this latter regard, gravitational lensing has been fundamental \cite{massey/2010,jung/2019,courbin/2000}.

\smallskip 
At least a portion of DM may be in the form of massive compact halo objects or MACHOs \cite{renault/1997,sackett/1993,wu/1994}. Those are massive baryonic matter objects that emit low or no electromagnetic radiation and habit galactic halos, and an example of them would be neutron stars. They can also bend light, causing gravitational microlensing effects, that have been detected for some time \cite{alcock/1997,gould/2004,thomas/2005,griest/1992,bennett/1995}. 

\smallskip

It should also be quoted that DM gravitational effects could be understood as purely geometrical effects of extended gravity theories \cite{capozziello/2013}. Rotation curves \cite{capozziello/2007,naik/2018,cardone/2010} and even structure formation \cite{dodelson/2006,acquaviva/2005,bebronne/2007,koyama/2006,pal/2006} have been explained through the extended gravity channel.

\smallskip

Here, in the present article, based on some of the several studies that empirically prove DM existence \cite{clowe/2006}, we are going to stick to the standard approach, considering DM exists and is non-baryonic.

\smallskip

The Bose-Einstein condensate is a possibility in the DM particle scenario \cite{Harko1,sikivie/2009,Harko2,ChavHarko,harko/2015},  and it was recently shown that could exist in space by the Cold Atom Laboratory orbiting Earth on board the International Space Station  \cite{aveline/2020}.

\smallskip

The weakly interacting massive particles (so-called WIMPs) \cite{pospelov/2008,chang/2014,graesser/2011} are among the best motivated DM particle candidates. 
WIMPS interact through a feeble new force and gravity as predicted by supersymmetry among other theories  \cite{roszkowski/2003,liu/2013}. If they were in thermal equilibrium in the early universe they annihilated with one another so that a predictable number of them remains today \cite{jungman/1996}. 

\smallskip

There may exist DM stars (DMSs) \cite{spolyar/2009,freese/2016} powered by WIMP DM annihilation \cite{Casanellas/2011,Lopes/2011}. In regions of high DM density, such as the Galactic center, the capture and annihilation of WIMP DM by stars has the potential to significantly alter their evolution \cite{scott/2009,Lopes/2014,Lopesa/2019,Lopesb/2019}. In Reference \cite{perez-garcia/2010} it was shown that WIMPs accreted onto neutron stars may provide a mechanism to seed strangelets in compact objects for WIMP masses above a few GeV. This effect may trigger a conversion of most of the star into a strange star.  Recall that neutron stars are pulsars, high-density stars with large rotation frequency rates located in the core of supernovae remnants \cite{lattimer/2004,lattimer/2001}. Some models predicted that strange stars could form inside these stars due to the brake neutrons into their constituent quarks \cite{alcock/1986}.  Due to a matter of stability, a portion ($\sim1/3$) of these quarks is converted to strange quarks and the resulting matter is known as strange quark matter.

\smallskip

Neutron stars are expected to efficiently capture WIMPs due to their strong gravitational field. The annihilation of DM in the center of these stars could lead to detectable effects on their surface temperature, specially if they are in the center of our Galaxy \cite{kouvaris/2010}. 

\smallskip

In \cite{kurita/2016}, Kurita and Nakano investigated the collapse of clusters of WIMPs in the core of Sun-like stars and the consequent possible formation of mini-black holes, which would generate gravitational wave emission. 

\smallskip

The aforementioned Bose-Einstein condensate has also been considered as the DM modeling for stars. On this regard, one can consult e.g. \cite{madarassy/2015,li/2012,DMS1,DMS2,DMS3,DMS4,lopes/2018,panotopoulos/2018}. In particular, in \cite{lopes/2018,panotopoulos/2018}, DMSs were investigated in the Starobinsky model of gravity \cite{starobinsky/1980}. It has been shown in \cite{panotopoulos/2018} that DMSs have smaller radius and are slightly more massive in Starobinsky gravity.

\smallskip

In the present article we will assume a boson star as our model for DMS. A wide variety of boson stars have been proposed and investigated in the literature \cite{jetzer/1992,schunck/2003,gleiser/1988,liddle/1992,colpi/1986} (for some recent references on this subject, one can check \cite{delgado/2020,li/2020,choptuik/2019,guerra/2019}). Our DMS will be modeled from the equation of state (EoS) proposed in \cite{colpi/1986} (check also \cite{maselli/2017}). It is interesting to mention that some proposals for detecting boson stars were reported in \cite{olivares/2020,torres/2000,dabrowski/2000,vincent/2016,bramante/2013,macedo/2013}.

\smallskip

The environment inside DMSs is expected to be extremely dense, specially when neutron star-like objects are under consideration. Under such conditions of extreme density, anisotropy is expected to appear \cite{mak/2003,dev/2003,raposo/2019,isayev/2017}.

\smallskip

Anisotropy in neutron stars has been investigated in the literature. The hydrostatic features were firstly approached in \cite{heintzmann/1975}, where it was shown that deviations from isotropy would entail changes in the star maximum mass. This approach was extended in \cite{hillebrandt/1976} to also cover the problem of stability under radial and non-radial pulsations. The effects of anisotropy on slowly rotating neutron stars was studied in \cite{silva/2015}. In \cite{folomeev/2018}, anisotropic neutron stars were also considered in the framework of Starobinsky gravity. Further studies of anisotropic neutron stars can be seen in References \cite{doneva/2012,torres-sanchez/2019,setiawan/2017,setiawan/2019}.

\smallskip

To the best knowledge of the present authors, anisotropy has not yet been considered in DMSs. Such an investigation is the main goal of the present article. The plan of our work is the following: in the next section we briefly summarize the
structure equations describing hydrostatic equilibrium of anisotropic stars. In Section III we present the
exact analytical solution and we show that it is well behaved and realistic within General Relativity. Finally, we finish our work in Section 4 with the concluding remarks.

%%%%%%%%%%%%%%%%%%%%%%%%%%%%%%%%%%%%%%%%%%%%%%%%%%%%
\section{Relativistic stars with anisotropic matter}
%%%%%%%%%%%%%%%%%%%%%%%%%%%%%%%%%%%%%%%%%%%%%%%%%%%%

Within General Relativity the starting point is Einstein's field equations 
\begin{equation}
\mathcal{G}_{\mu \nu} = \mathcal{R}_{\mu \nu}-\frac{1}{2} \mathcal{R} g_{\mu \nu}  = 8 \pi T_{\mu \nu}.
\end{equation}
In (1), $\mathcal{G}_{\mu\nu}$ is the Einstein tensor, $\mathcal{R}_{\mu\nu}$ is the Ricci tensor, $\mathcal{R}$ is the Ricci scalar, $g_{\mu\nu}$ is the metric tensor, we set Newton's constant $G$ and the speed of light, $c$, to $1$, while for anisotropic matter the stress-energy tensor, $T_{\mu\nu}$, has the form
\begin{equation}
T_\nu ^\mu = \text{Diag}(-\rho, p_r, p_t, p_t),
\end{equation}
with $\rho$ being the energy density, $p_r$ the radial pressure and $p_t$ the tangential pressure.

In order to find interior solutions describing hydrostatic equilibrium of relativistic stars, we integrate the structure equations including the presence of a non-vanishing anisotropic factor \cite{anisotropies,PR2}:
\begin{eqnarray}
m'(r) & = & 4 \pi r^2 \rho(r), \\
\nu'(r) & = & 2 \frac{m(r)+4 \pi r^3 p_r(r)}{r^2 [1-2m(r)/r]}, \\
p_r'(r) & = & - [\rho(r) + p_r(r)] \: \frac{m(r)+4 \pi r^3 p(r)}{r^2 [1-2m(r)/r]} + \frac{2 \Delta}{r}, 
\end{eqnarray}
where $m(r)$ and $\nu(r)$ are the components of the metric tensor assuming static, spherically symmetric solutions in Schwarzschild-like coordinates, $(t,r,\theta,\phi)$, 
\begin{equation}
ds^2 = -e^{\nu} dt^2 + \frac{1}{1-2m(r)/r} dr^2 + r^2 (d \mathrm{\theta^2} + \mathrm{sin^2 \theta \: d \phi^2}),
\end{equation}
and $\Delta \equiv p_t - p_r$ is the anisotropic factor. All quantities depend on the radial coordinate $r$ only,
and a prime denotes differentiation with respect to $r$. Clearly, setting $\Delta=0$ we recover the usual Tolman-Oppenheimer-Volkoff equations \cite{Tolman,OV} for isotropic matter.

\smallskip

Moreover we impose at the center of the star, $r=0$, the following initial conditions
\begin{eqnarray}
m(0) & = & 0,  \\
p(0) & = & p_c,
\end{eqnarray}
with $p_c$ being the central pressure. Upon matching with the exterior vacuum solution ($T_{\mu \nu}=0$, Schwarzschild geometry) at the surface of the star, $r=R$, the following boundary conditions must be satisfied
\begin{eqnarray}
p(R) & = & 0,  \\
m(R) & = & M, \\
e^{\nu(R)} & = & 1 - \frac{2M}{R},
\end{eqnarray}
with $R$ being the radius of the star, and $M$ being its mass.

%%%%%%%%%%%%%%%%%%%%%%%%%%%%%%%%%%%%%%%%%%%%%%%%%%%%%%%%%%%%%%%%%%%%%%%%%
\section{{\bf Anisotropic} Dark matter stars: Exact analytical solution}
%%%%%%%%%%%%%%%%%%%%%%%%%%%%%%%%%%%%%%%%%%%%%%%%%%%%%%%%%%%%%%%%%%%%%%%%%

Boson stars are self-gravitating clumps made of either spin-zero fields, called scalar boson
stars \cite{schunck/2003} or vector bosons, called Proca stars \cite{Proca1,Proca2}. The maximum mass for 
scalar boson stars in non-interacting systems was found in \cite{BS1,BS2}, while in \cite{colpi/1986,BS3} it 
was pointed out that self-interactions can cause significant changes.

A complex scalar field, $\Phi$, minimally coupled to gravity is described by the Einstein-Klein-Gordon action \cite{Liebling}
\begin{eqnarray}
S & = & \int d^4 x \sqrt{-g} \left( \frac{R}{16 \pi} + \mathcal{L}_M  \right) \\
\mathcal{L}_M & = & - g^{\mu\nu} \partial_\mu \Phi \partial_\nu \Phi^* - V(|\Phi|)
\end{eqnarray}
where $g$ is the metric determinant, $\mathcal{L}_M$ is the matter lagrangian and $V$ is the self-interaction scalar potential.

For static spherically symmetric solutions we make for the scalar field the ansatz \cite{Liebling}
\begin{equation}
\Phi(r,t) = \phi(r) exp({-i \omega t}),
\end{equation}
where the oscillation frequency $\omega$ is a real parameter. 

Although the scalar field itself depends on time, its 
stress-energy tensor is time independent and the Einstein's field equations take the usual form for a fluid, for 
which the energy density is computed to be \cite{Cardoso1,Cardoso2}
\begin{equation}
\rho = \omega^2 e^{-\nu} \phi^2 + e^{-\lambda} \phi'^2 + V(\phi),
\end{equation}
while the radial and tangential pressures are found to be \cite{Cardoso1,Cardoso2}
\begin{eqnarray}
p_r & = & \omega^2 e^{-\nu} \phi^2 + e^{-\lambda} \phi'^2 - V(\phi), \\
p_t & = & \omega^2 e^{-\nu} \phi^2 - e^{-\lambda} \phi'^2 - V(\phi).
\end{eqnarray}

Clearly, a boson star is anisotropic since the two pressures are different. Under certain conditions, however, the anisotropy may be ignored and the system can be treated as an isotropic object. A concrete model of the form
\begin{equation}
V(|\Phi|) = m^2 |\Phi|^2 + \frac{\lambda}{2} |\Phi|^4,
\end{equation}
with $m$ being the mass of the scalar field and $\lambda$ being the self-interaction coupling constant,
was studied e.g. in \cite{maselli/2017}, in which the authors considered the following EoS \cite{colpi/1986}:
\begin{equation}
p_r = \frac{\rho_0}{3} \left( \sqrt{1+\frac{\rho}{\rho_0}}  - 1 \right)^2,
\end{equation}
where $\rho_0$ is a constant given by
\begin{equation}
\rho_0 = \frac{m^4}{3 \lambda}.
\end{equation}
This EoS describes the boson stars that are approximately isotropic provided that the condition
\begin{equation}
\frac{\lambda}{4 \pi} \gg m^2
\end{equation}
holds \cite{maselli/2017}. 

In the two extreme limits we recover the well-known results
\begin{equation}
p_r \approx  \frac{\rho^2}{12 \rho_0}, \ \quad \rho \ll \rho_0,
\end{equation}
for diluted stars \cite{ChavHarko}, and 
\begin{equation}
p_r \approx \frac{\rho}{3}, \ \quad \rho \gg \rho_0
\end{equation}
in the ultra relativistic limit.

In the first extreme limit, any model, irrespectively of the form of the potential, will be described by the 
same polytropic EoS, with index $n=1$ and $\gamma=2$. In the present work we propose to investigate the properties of
relativistic stars made of anisotropic exotic matter characterized by the polytropic EoS 
\begin{equation}
p_r = K \rho^2, \; \; \; \; \; K=z/B
\end{equation}
where $z$ is a dimensionless number while $B$ has dimension of pressure and it is of the order of the energy density of neutron stars and quark stars, $B \simeq (150~MeV)^4$.

In the case of stars with anisotropic matter there are five unknown quantities in total and only three differential equations. Therefore, we are free to impose two conditions. Given the EoS, the simplest thing to do is to assume a certain profile for the energy density. In the following we shall consider the ansatz 
\begin{equation}
\rho(r) = \rho_c \left( 1 - \frac{r^2}{R^2} \right),
\end{equation}
which ensures that the energy density starts from a finite value at the origin, which is the central value $\rho_c$, and it monotonically decreases with $r$, until it vanishes at the surface of the star. 

Now, all the other quantities may be computed one by one using the structure equations and the EoS. In particular, the radial pressure is immediately computed
making use of the EoS, while the mass function is computed using the {\it tt} component of the field equations, and it is given by
\begin{equation}
m(r) = 4 \pi \int_0^r dx x^2 \rho(x) = 4 \pi \rho_c r^3 \left( \frac{1}{3} - \frac{r^2}{5R^2} \right).
\end{equation}

The temporal metric component $\nu$ is computed making use of the radial field equation as follows
\begin{equation}
\nu(r) = log(1-2 M/R) - 2 \int_R^r dx \frac{m(x)+4 \pi x^3 p_r(x)}{x^2 [1-2m(x)/x]}.
\end{equation}

Finally, the anisotropic factor is computed making use of the conservation of energy
\begin{equation}
\Delta(r) = \frac{r}{2}\left\{[p_r(r)+\rho(r)]\frac{m(r)+4 \pi r^3 p_r(r)}{r^2 [1-2m(r)/r)} + p_r'(r)]\right\},
\end{equation}
while the tangential pressure is computed from $p_t(r) = p_r(r) + \Delta(r)$.

Next we shall investigate the behavior as well as the viability of the solutions we just found.

%%%%%%%%%%%%%%%%%%%%%%%%%%%%%%%%%%%%%%%%%%%%%%%%%%%%%%%%%%
\subsection{Causality, stability and energy conditions}
%%%%%%%%%%%%%%%%%%%%%%%%%%%%%%%%%%%%%%%%%%%%%%%%%%%%%%%%%%

The radial and tangential speeds of sound, defined by 
\begin{eqnarray}
c_r^2 & \equiv & \frac{dp_r}{d \rho}, \\
c_t^2 & \equiv & \frac{dp_t}{d \rho},
\end{eqnarray}
should take values in the interval $0 < c_{r,t}^2 < 1$ throughout the stars, so that causality is not violated. 

Moreover, Bondi suggested that for a stable Newtonian sphere the radial adiabatic index, defined by
\begin{equation}
\Gamma \equiv c_r^2 \left( 1 + \frac{\rho}{p_r} \right),
\end{equation}
should be larger than 4/3 \cite{bondi64}. 

Finally, the solutions obtained here should be able to describe realistic astrophysical configurations. Therefore, as a further check we investigate if the energy conditions are fulfilled or not. To that end, the conditions \cite{Ref_Extra_1,Ref_Extra_2,Ref_Extra_3,ultimo,PR3}
\begin{equation}
\rho \geq 0\,,
\end{equation}
\begin{equation}
\rho + p_{r,t}  \geq  0\,,
\end{equation}
\begin{equation}
\rho - p_{r,t}  \geq  0\,,
\end{equation}
\begin{equation}
E_+ \equiv \rho + p_r + 2 p_t \geq 0\,,
\end{equation}
\begin{equation}
E_- \equiv \rho - p_r - 2 p_t \geq 0\,,
\end{equation}
are investigated.

Our main numerical results are summarized in Figures \ref{fig:1}, \ref{fig:2} and \ref{fig:3} below, assuming the following numerical values for $z$, $B$ and $\rho_c$:
\begin{eqnarray}
B & = & 2 \times 10^{-80}~m_{pl}^4, \\
\rho_c & = & 14 \: B, \\
z & = & 0.009,
\end{eqnarray}
with $m_{pl}$ being the Planck mass.
(27) and (28) correspond to a star with the following properties 
\begin{eqnarray}
R & = & 12~km, \\
M & = & 2.08~M_{\odot}, \\
C & = & 0.258,
\end{eqnarray}
where $C=M/R$ is the compactness factor of the star.

%%%%%%%%%%%%%%%%%%%%%%%%%%%%FIGURES%%%%%%%%%%%%%%%%%%%%%%%%%%%%%%%%%%%%%%

\begin{figure}[ht!]
\centering
\includegraphics[width=\linewidth]{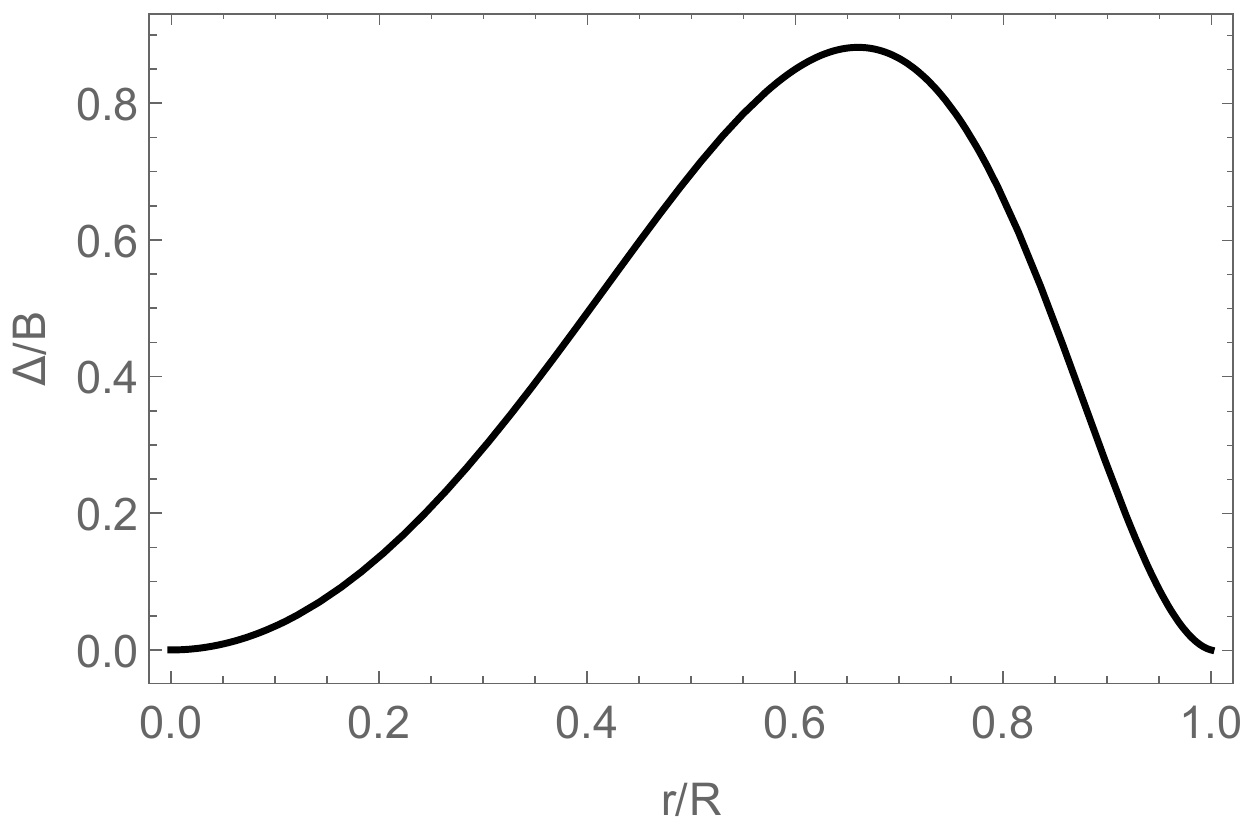} 
\caption{
Normalized anisotropic factor, $\Delta/B$, versus normalized radial coordinate $r/R$.
}
\label{fig:1} 	
\end{figure}

%%%%%%%%%%%%%%%%%%%%%%%%%%%

\begin{figure}[ht!]
\centering
\includegraphics[width=\linewidth]{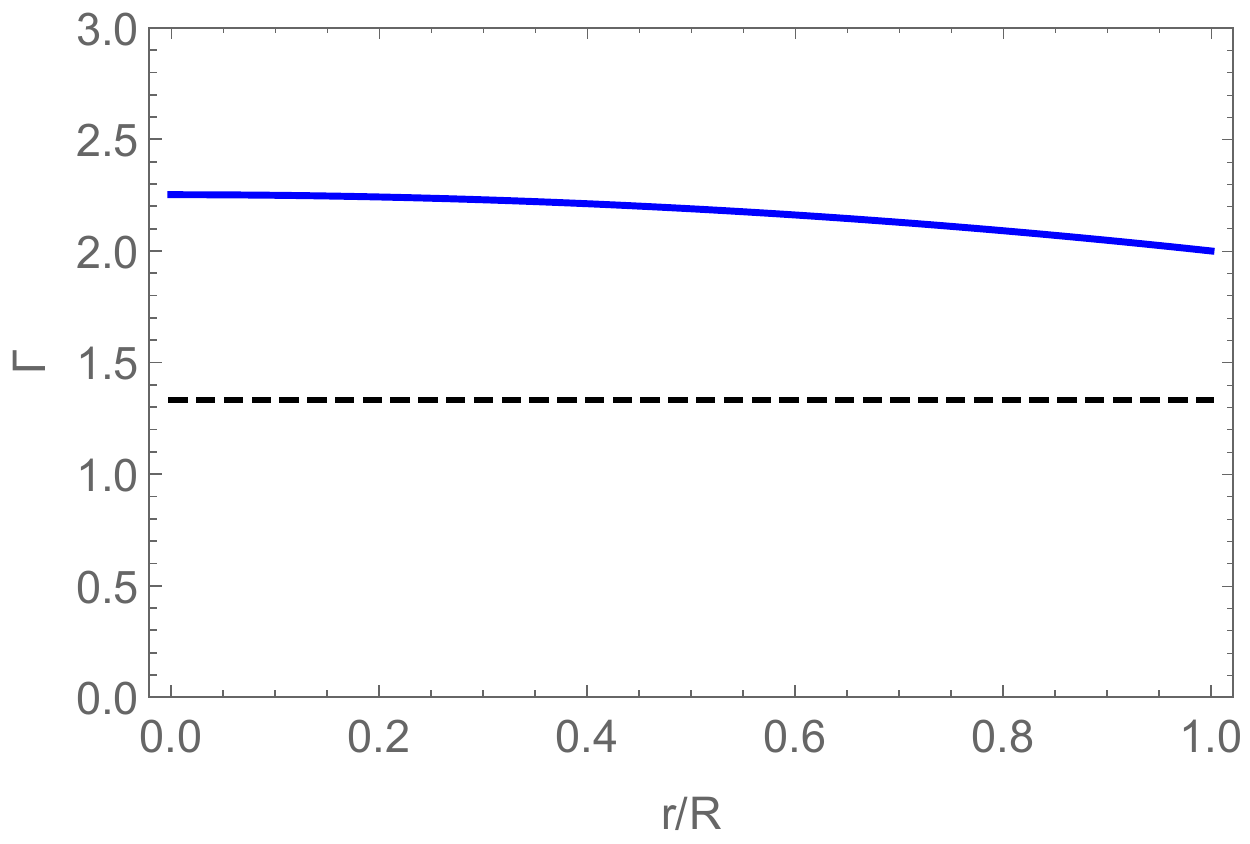} 
\caption{
Relativistic adiabatic index, $\Gamma$, versus normalized radial coordinate $r/R$. The horizontal line corresponds to
the Newtonian limit of $4/3$.
}
\label{fig:2} 	
\end{figure}

%%%%%%%%%%%%%%%%%%%%%%%%%%%%%

\begin{figure}[ht!]
\centering
\includegraphics[width=\linewidth]{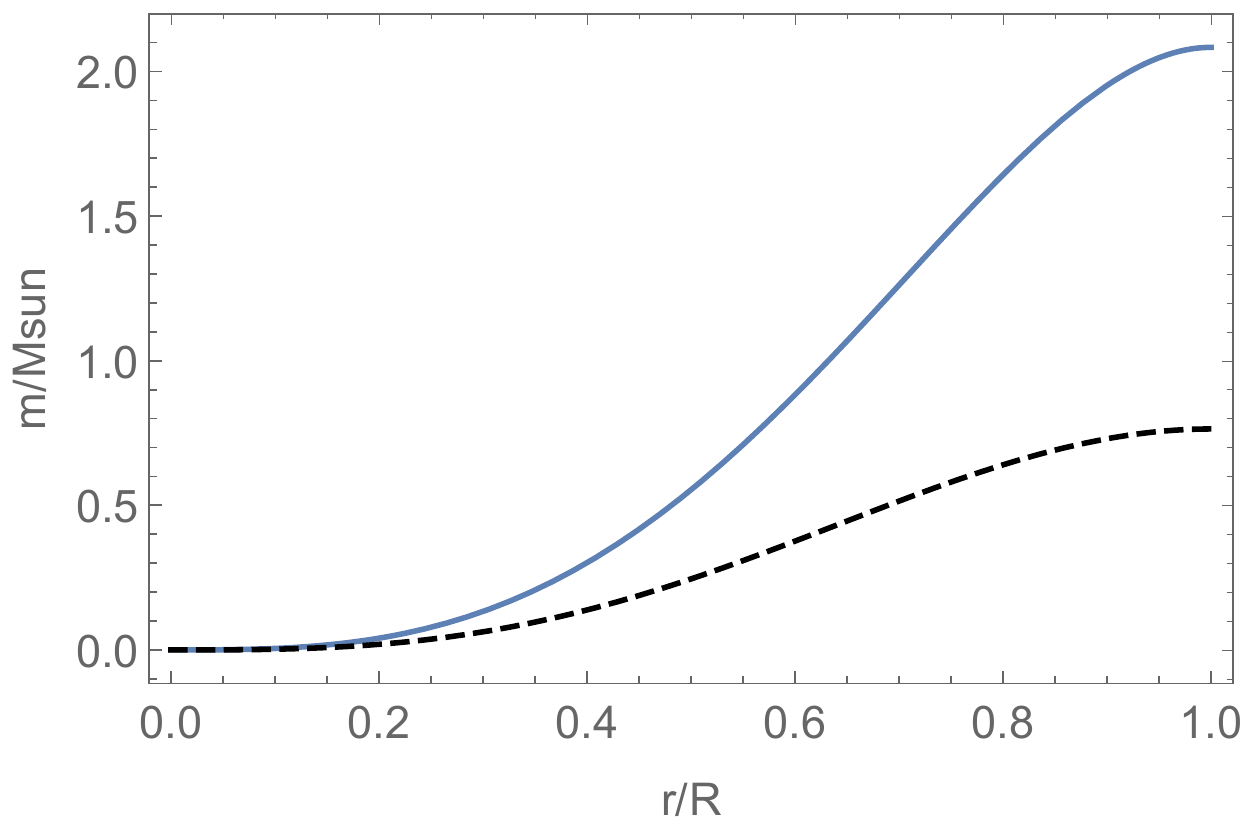} 
\caption{
Mass function (in solar masses) versus normalized radial coordinate $r/R$. The dashed curve corresponds to isotropic stars.
}
\label{fig:3} 	
\end{figure}

%%%%%%%%%%%%%%%%%%%%%%%%%

\begin{figure}[ht!]
\centering
\includegraphics[width=\linewidth]{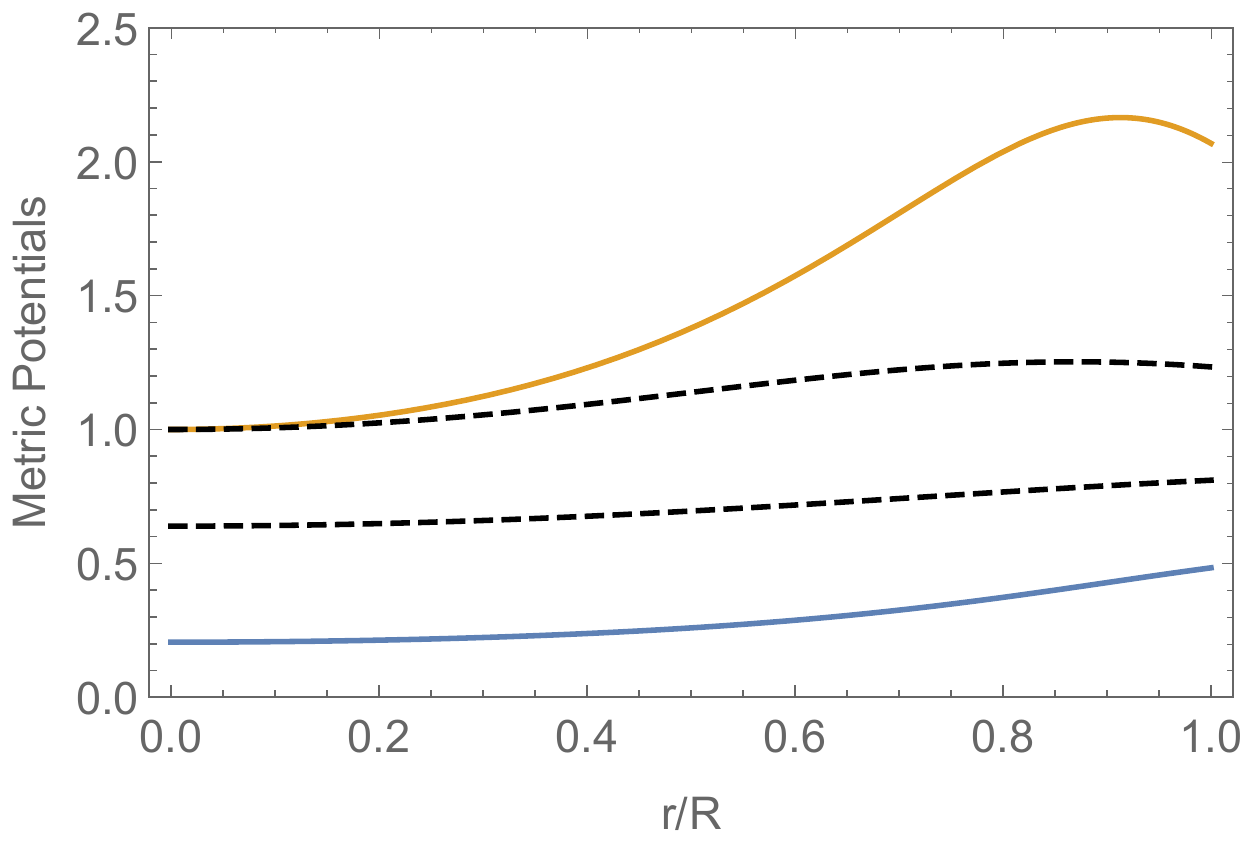} 
\caption{The two metric components, $e^\nu$ (lower curve) and $1/(1-2m/r)$ (upper curve) versus normalized radial coordinate $r/R$.
The dashed curves correspond to isotropic stars.}
\label{fig:4} 	
\end{figure}

%%%%%%%%%%%%%%%%%%%%%%%%%%%%%%%%%%%%

\begin{figure}[ht!]
\centering
\includegraphics[width=\linewidth]{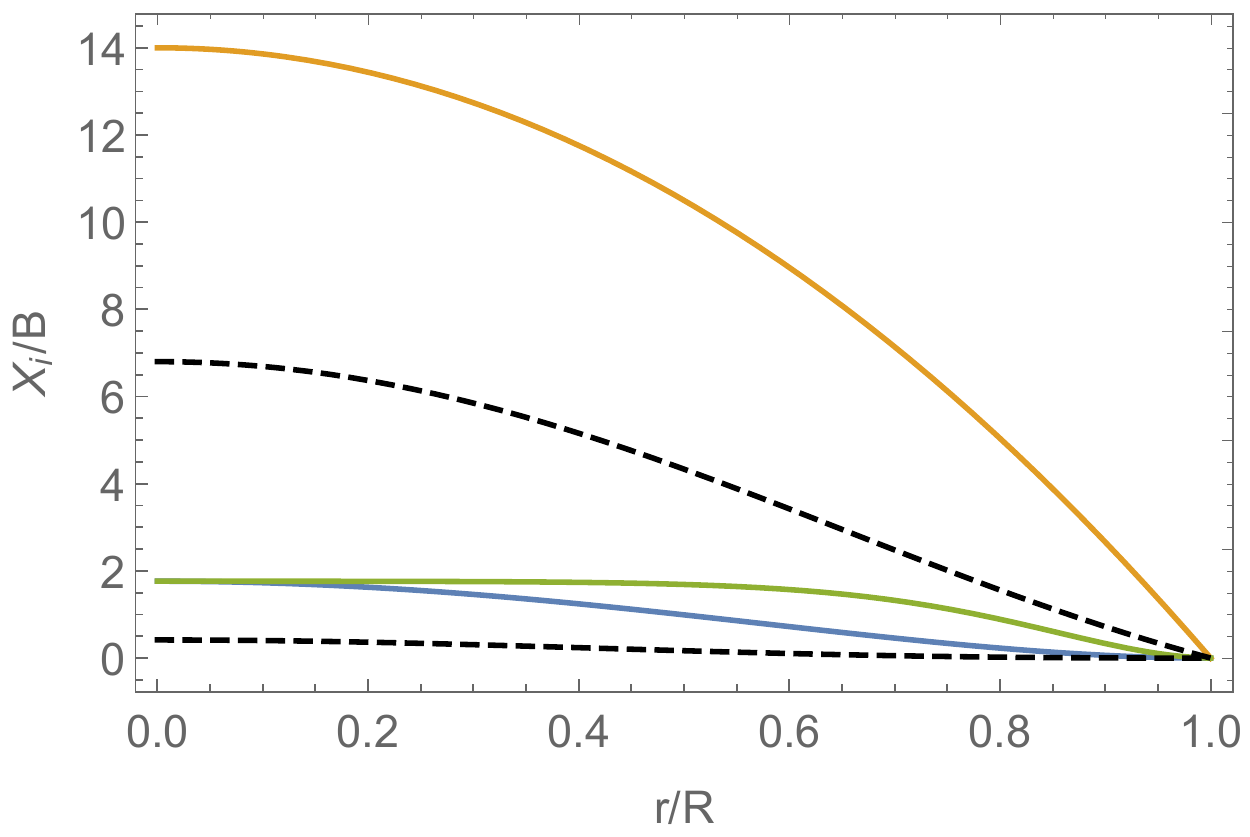} 
\caption{
Energy density $\rho/B$ (orange curve) radial pressure $p_r/B$ (blue curve) and tangential pressure $p_t/B$ (green curve)
versus $r/R$. The dashed curves correspond to isotropic stars, where energy density lies above pressure.
}
\label{fig:5} 	
\end{figure}

%%%%%%%%%%%%%%%%%%%%%%%%%%%%%%%%%%%%

\begin{figure}[ht!]
\centering
\includegraphics[width=\linewidth]{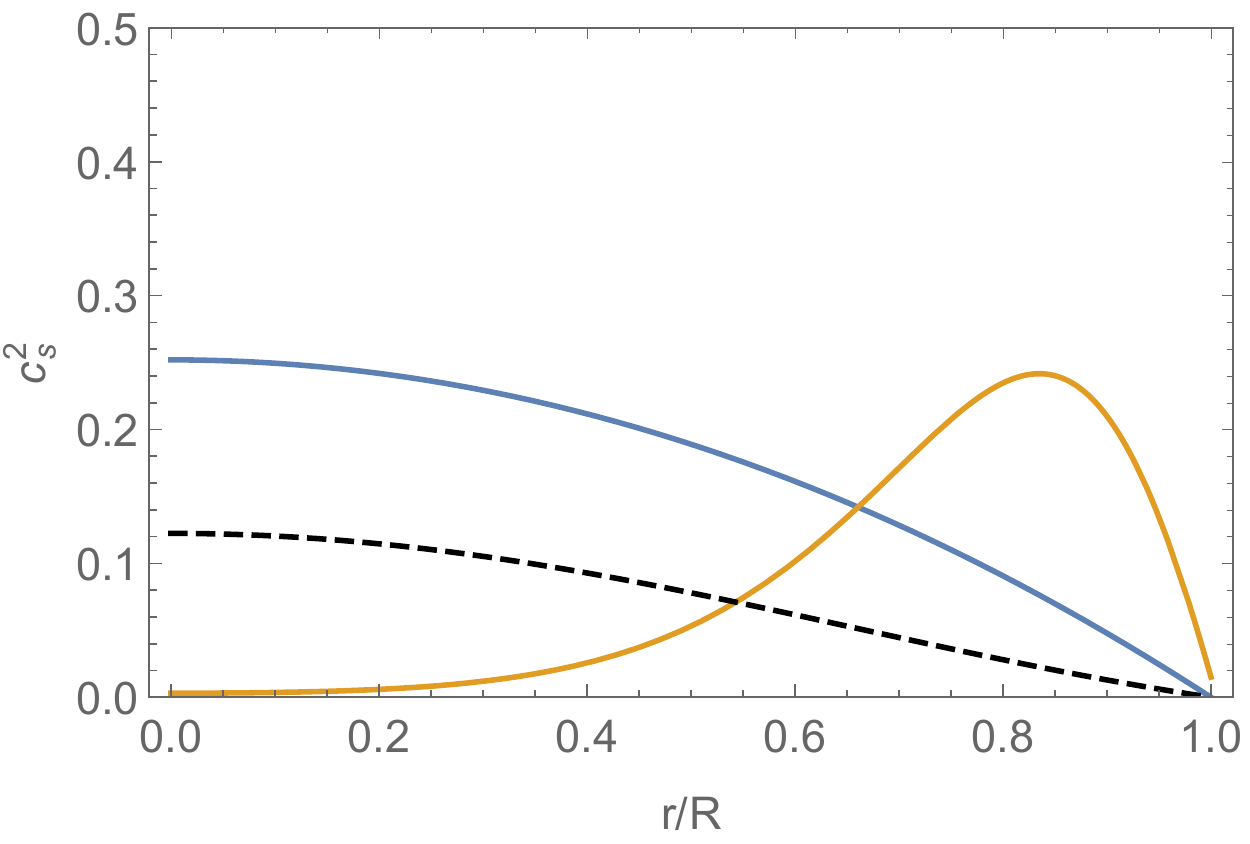} 
\caption{
Radial (blue curve) and tangential (orange curve) sound speeds, $c_r^2, c_t^2$, versus normalized radial coordinate $r/R$. The dashed curve corresponds to isotropic stars.
}
\label{fig:6} 	
\end{figure}

%%%%%%%%%%%%%%%%%%%%%%%%%%%%%FIGURES%%%%%%%%%%%%%%%%%%%%%%%%%%%%%%%%%%%%%%%

In particular, Fig.~\ref{fig:1} shows the normalized anisotropic factor $\Delta(r)/B$ versus $r/R$.
It vanishes both at the center and at the surface of the star and it is positive throughout the object.
The adiabatic index $\Gamma$ versus $r/R$ is shown in Fig.~\ref{fig:2}, where the Newtonian limit of $4/3$
is shown as well. 

In the Figures 3-6 a comparison is made between stars with anisotropic matter
and their isotropic counterparts with the same EoS and the same radius. In particular, in Fig.~\ref{fig:3} 
we show the mass functions versus $r/R$, while Fig.~\ref{fig:4} shows the two metric potentials versus $r/R$. 
Finally, Fig.~\ref{fig:5} shows normalized energy density and pressures versus $r/R$, while in Fig.~\ref{fig:6} 
we show the speeds of sound, both radial (blue curve) and tangential (orange curve), versus $r/R$.

Clearly, causality is not violated as both sound speeds take values in the range $(0,1)$ throughout the star. 
Moreover, the condition $\Gamma > 4/3$ is satisfied as well. Finally, since both pressures are positive and lower 
than the energy density, all energy conditions are fulfilled. 

%%%%%%%%%%%%%%%%%%%%%%%
\section{Conclusions}
%%%%%%%%%%%%%%%%%%%%%%%

In summary, in the present work we have studied exotic stars with anisotropic matter within General Relativity. 
We have investigated in detail the properties of dark matter-type configurations, taking into account the presence of anisotropies. Exact analytic expressions for all the quantities of interest, such as mass function, anisotropic factor, relativistic index, speed of sound etc, have been found. Causality, stability criteria and energy conditions are also discussed. It is found that the solutions obtained here are well-behaved solutions capable of describing realistic astrophysical configurations. Finally, a direct comparison with their isotropic counterparts was made as well.

%%%%%%%%%%%%%%%%%%%%%%%%%%%%%%%%%%%%%%%%%%%%%%%%%%%%%%

\section*{Acknowledgments} 

PHRSM thanks CAPES for financial support. The authors G.~Panotopoulos and I.~Lopes 
thank the Funda\c c\~ao para a Ci\^encia e Tecnologia (FCT), Portugal, for the 
financial support to the Center for Astrophysics and Gravitation-CENTRA, Instituto 
Superior T\'ecnico, Universidade de Lisboa, through the Project No.~UIDB/00099/2020 and grant No. PTDC/FIS-AST/28920/2017.
%%%%%%%%%%%%%%%%%%%%%%%%%%%%%%%%%%%%%%%%%%%%%%%%%%%%%%%

\end{document}